
\input harvmac
\def\m{\mu}
\def\n{\nu}
\def\p{\phi}
\def\pa{\partial}
\def\d{\delta}
\def\l{\lambda}
\def\s{\sigma}
\def\r{\rho}
\def\b{\beta}
\Title{USITP-91-19}
{{\vbox{\centerline{ The Path Integral for a Particle
in  Curved Spaces}\bigskip
 \centerline {
and Weyl Anomalies}}}}

\centerline{Fiorenzo Bastianelli\footnote{$^\dagger$}{e-mail:
fiorenzo@vana.physto.se}}
\bigskip\centerline
{\it  Institute for Theoretical Physics}
\centerline {\it University of Stockholm}
\centerline {\it Vanadisv\"agen 9 }
\centerline {\it S-113 46 Stockholm, Sweden}
\vskip .2in

\noindent
The computation of anomalies in quantum field theory
may be carried out by evaluating  path integral Jacobians, as
first shown by Fujikawa.
The evaluation of these Jacobians can be cast in the form of
a quantum mechanical problem, whose solution has a
path  integral representation. For the case of Weyl anomalies, also
called trace anomalies,
one is immediately led to study the path integral for a particle moving
in curved spaces. We analyze the latter
in a manifestly covariant way and by  making use of ghost fields.
The introduction of the  ghost fields allows us  to represent
the path integral measure in a form suitable for performing
the perturbative
expansion.
We employ our method to compute the Hamiltonian associated with
the evolution kernel
given by the path integral with fixed boundary conditions,
and use this result to evaluate the trace needed
in field theoretic computation of Weyl anomalies in two dimensions.

\Date{11/91}

\newsec{Introduction}

Anomalies arise in quantum field theory whenever the symmetries of a
classical system cannot be all preserved by the quantization procedure
\ref\ABJ{S.L. Adler, {\sl Phys. Rev.} {\bf 117} (1969) 2426\semi
J. Bell and R. Jackiw, {\sl Nuovo Cimento} {\bf A47} (1969) 47.}.
In a path integral context, this happens when the path integral measure
is not invariant under certain symmetries. When deriving Ward identities
for these  particular symmetries by performing a change of path integration
variables, a Jacobian appears. This Jacobian is identified as the anomaly
\ref\fuji{K. Fujikawa, {\sl Phys. Rev. Lett.} {\bf 44} (1980) 1733;
{\sl Phys. Rev.}
{\bf D21} (1980) 2848; {\bf D23} (1981) 2262.}.
For change of variables given by  infinitesimal symmetry transformations,
the Jacobian differs from unity by a trace which may
be evaluated in several ways.
One of the most elegant ways, which  goes back to Schwinger
\ref\Schwinger{J. Schwinger, {\sl Phys. Rev.} {\bf 82} (1951) 664.},
 consists in representing the operator we need to compute the trace of
by a quantum mechanical operator acting in a fictitious  (non-physical)
Hilbert space. Using then the reformulation of quantum mechanics due to
Feynman
\ref\Feynman{R.P. Feynman, {\sl Rev. Mod. Phys.} {\bf 20} (1948) 367\semi
R.P. Feynman and A.R. Hibbs,{\sl ``Quantum Mechanics and Path Integrals''},
McGraw-Hill, New York,1965.},
the trace computation is recast as a path integral with specific
boundary conditions and  evaluated.
Such a program was beautifully carried out in the computation
of chiral anomalies
\ref\ag{L. Alvarez-Gaum\'e, {\sl Comm. Math. Phys.} {\bf 90}
(1983)  161\semi
L. Alvarez-Gaum\'e and E. Witten, {\sl Nucl. Phys.} {\bf B234} (1984) 269\semi
D. Friedan and P. Windey, {\sl Nucl. Phys.} {\bf B234\ [FS11]} (1984) 395.},
where
 supersymmetric quantum mechanics was used.
The topological characterization of chiral anomalies is  responsible
for the relative easiness in computing the corresponding
path integral. In fact,
because of the topological character, any
approximation may be used to evaluate such a path integral
and a semiclassical approximation
gives the full result.
However, the method is general and one may try to use it to
compute other types of anomalies. In the present article, we look
at the case of Weyl anomalies, also called trace anomalies \ref\?{
D.M. Capper and M.J. Duff, {\sl Nuovo Cimento} {\bf 23A} (1974) 173.}.
Weyl transformations are  defined by  arbitrary rescalings of
the metric tensor. A field theory invariant under such
 rescalings (and corresponding rescalings of the dynamical fields)
has necessarily an energy-momentum tensor with a vanishing trace.
An anomaly in the Weyl symmetry is then seen as  a non-vanishing trace in
the energy-momentum tensor arising quantum mechanically
(hence the alternative name of trace anomalies).
In the following, we will
consider the special case of a scalar field coupled
to gravity in a Weyl invariant way. After having identified
the trace giving the Weyl anomaly, we look for a quantum
mechanical representation
of such a trace. The relevant quantum mechanical system turns out to be
that of a particle moving in a curved space. The path integral for
such a system has been  a favorite topic of study over the years
\ref\manypeople{B.S. De Witt, {\sl Rev. Mod. Phys.} {\bf 29} (1957)
377\semi
H. Dekker, {\sl Physica} {\bf 103A} (1980) 586.},\ref\Schul{ L. Schulman,
{\sl ``Techniques
and Applications of Path Integration''}, John Wiley and Sons, New York,
 1981, and references therein.}.
However,  we try to do better of what we could find in the literature
and develop a method for computing perturbatively the path integral for the
particle in a manifestly geometrical way (i.e. using Riemann normal
coordinates) and making use of ghost fields. These ghost fields are
essential to exponentiate a certain path integral
Jacobian, thereby leaving a translational
invariant measure suitable for performing a perturbative expansion.
The Weyl anomaly does not seem to have a topological interpretation and this
may be the reason why a semiclassical approximation is not enough to get the
complete
 result. Our findings are that the Weyl anomaly for the Klein Gordon field
in $2k$ dimensions (Weyl anomalies are present only in even dimensions)
is obtained by an $k+1$ loop computation in the corresponding
quantum mechanical system.

The paper is structured as follows. In section 2
we  describe some generalities
on Weyl transformations and identify the ``Fujikawa''
trace which gives
the anomaly for the Klein-Gordon field in arbitrary dimensions.
In section 3 we discuss
the quantum mechanics of a particle  in a curved space. We define its
path integral in a manifestly covariant manner and by making use of ghost
fields. In section 4
we evaluate our reformulation of
the path integral in the two loop approximation
and obtain the  Hamiltonian associated with the evolution kernel. This gives
enough information for obtaining the Weyl anomaly in two dimensions.
Eventually, we state our conclusions in section 5, where we
 comment on the Weyl anomaly in higher dimensions
as well as on the
connection of our computation with the Schwinger-DeWitt method for evaluating
the Klein-Gordon propagator in curved spaces
\ref\dewitt{B.S. De Witt in {\sl ``Relativity, Groups and Topology''},
 edited by
B.S. De Witt and C. De Witt, Gordon Breach, New York,  1964;
 {\sl ``Relativity, Groups and Topology II''}, edited by
B.S. De Witt and R. Stora, North Holland, Amsterdam, 1984.}.

\newsec{Generalities on Weyl transformations}
Let's consider a classical field theory in a curved space defined by an
action functional $S[\p,g_{\m\n}]$, where the dynamical fields
 are collectively
denoted by $\p$ and where $g_{\m\n}$ is the metric tensor of
the manifold, considered as a background.
 A Weyl transformation is defined as an arbitrary rescaling
of the metric
\eqn\res{g_{\m\n}'(x) = \Omega(x) g_{\m\n}(x)}
The theory defined by the action
$S[\phi,g_{\m\n}]$ is Weyl invariant if one can
define transformation rules on the dynamical fields $\phi$ which, together
with \res,  leave the action  unchanged.
In those cases where Weyl invariance can be achieved,
the energy-momentum
tensor, defined as $T^{\m\n} = {2 \over \sqrt g}
{\d S \over \d g_{\m\n}}$,
is necessarily traceless. This follows by considering
an infinitesimal Weyl transformation, which being a symmetry
leaves the action unchanged
\eqn\varia{ \d S[\phi, g_{\m\n}] = \int d^nx\ \biggl
({\d S \over {\d g_{\m\n}(x)}}
\d g_{\m\n}(x) + {\d S \over{ \d \p (x)}} \d \p (x) \biggr ) = 0}
Infinitesimally $\d g_{\m\n}(x) = \s(x) g_{\m\n}(x)$ with $\s(x)$
an arbitrary local function
and it follows from \varia\
that, on the shell of the $\p$  equations of motion
(${\d S \over \d \p}=0 $), the trace of the energy momentum vanishes
\eqn\traccia{ T\equiv g^{\m\n} T_{\m\n} =0}
Strictly speaking, this is not a true symmetry as long as the
metric is not dynamical. In fact, changing the metric is like changing
 coupling constants, which are defining parameters of the theory.
Truly dynamical symmetries are obtained by considering
the conformal group. This is defined as the subgroup of coordinate
transformations which change the metric only up to a scale factor
$f(x)$
\eqn\conf{ g_{\m\n}'(x') = g_{\r\s}(x)^{\ }{\pa x^\r \over {\pa x'{}^\m}}
{\pa x^\s \over{ \pa x'{}^\n}} = f(x) g_{\m\n}(x) }
Then, by fine tuning $\Omega (x) = (f (x))^{-1}$ in \res\
to compensate  for the change in \conf,
one is able to
construct transformation rules which leave  $g_{\m\n}$ inert,
while only the dynamical fields $\p$ get transformed.
In the case of flat space-time of signature $(q,p)$, one obtains
the usual conformal group,
locally isomorphic to $ SO(p+1,q+1) $\footnote{$^\dagger$}{Except in one
and two dimensions, where the conformal group is infinite dimensional.}.

A typical example of a Weyl invariant system is given by the massless
Klein Gordon field  with action
\eqn\KG{ S_{_{KG}} = \int d^n x {\sqrt g}\ {1\over 2} ( g^{\m\n}
\pa_\m \p \pa_\n \p - \xi R \p^2)}
where $n$ is the space-time dimensions, $\xi= { n-2 \over 4(n-1)}$
and $R$ is the curvature scalar. Our conventions for the curvature
tensor and the various related quantities
are encoded in the following relations:
$ [\nabla_\m, \nabla_\n] V^\r = R_{\m\n}{}^\r{}_\s V^\s$,
$R_{\m\n} = R_{\m\s}{}^\s{}_\n$ and  $R=g^{\m\n} R_{\m\n}$, where
$V^\r$ is an arbitrary vector
and $\nabla_\m $ is the usual covariant derivative.
 The Weyl transformation rules which leave the action \KG\
invariant are defined as follows
\eqn\a{\eqalign{ &{g_{\m\n}}' = \Omega g_{\m\n} \cr
&{\p}' = \Omega^{{
{2-n} \over 4}} \p \cr}}
and infinitesimally they read as
\eqn\W{\eqalign{ &\d g_{\m\n} = \s g_{\m\n} \cr
&\d \p = { 1
\over 4}(2 -n)  \s \p \cr}}
where $\s = \ln \Omega$ is taken to be small.
The coupling to the scalar curvature $R$ is necessary  to insure invariance
under local Weyl transformations.
For $\xi \neq {n-2 \over {4 (n-1)}}$ only a  few rigid Weyl
symmetries survive, and precisely those for which $\s$ is harmonic (i.e.
satisfies $\nabla^2 \s =0$). An example of this case is given by
the transformation with $\s$
constant,
which  implies a less restrictive condition then \traccia, namely
 the vanishing of the trace of the energy-momentum
tensor integrated over space-time.

To analyze  the quantum theory, we consider the Euclidean path integral
\eqn\Z{Z[g_{\m\n}] = \int {\cal D} \p \  \exp \biggl [
{-{1\over \hbar}S_{_{KG}}} \biggr ]}
The Ward identities which express the Weyl invariance of the classical theory
can be derived by performing a dummy change of
path integration variables
$\p \rightarrow \p' =\p +\d \p $ in \Z\ with $\d \p$ given in \W,
and  using the invariance of $S_{_{KG}}$. One obtains
\eqn\wi{\int d^n x {\sqrt g} \ \langle T \rangle \s = -2\hbar {\rm Tr}
 \biggl (
{\pa \d \p \over \pa \p }\biggl )}
where $\langle \cdots \rangle$ denotes the vacuum expectation value computed
with \Z.
The trace in the right hand side of \wi, if non-vanishing, gives the anomaly.
However, such a  trace is still a formal expression because the
operator $t(x,y) \equiv {\pa \d \p(x) \over \pa \p (y)}$
is infinite dimensional and one must discuss the necessary regularization
to define its trace.
This is achieved by using a negative definite operator ${\cal R}$
and defining the regularized trace by \fuji\
\eqn\regul{ {\rm Tr }
(t) = \lim_{m \to \infty} {\rm Tr} ( t e^{{{\cal R} \over m^2}})}
The effect of $ e^{{{\cal R}\over m^2}}$ is to exponentially damp the
contribution of the higher frequency modes, making the trace convergent.
Choices of ${\cal R}$ are not arbitrary if one wants to obtain
a consistent anomaly \ref\zum{W.A. Bardeen and B. Zumino,
{\sl Nucl. Phys.} {\bf B244} (1984) 421.}.
A general method \ref\peter{
A. Diaz, W. Troost, P. van Nieuwenhuizen and A. Van Proeyen,
{\sl Int. J. Mod. Phys.} {\bf A4} (1989) 3959\semi
M. Hatsuda, W. Troost, P. van Nieuwenhuizen and A. Van Proeyen,
{\sl Nucl. Phys.} {\bf B335} (1990) 166.}
to identify such a regulator is to appeal to
a Pauli-Villars regularization, which guarantees the consistency
of the anomaly. For further details on this procedure, we refer
directly to \peter.
For our purposes, we use
the Fujikawa variables  $ {\tilde \p } \equiv g^{1\over 4} \p $
to derive the Ward identity in \wi, so that
one is assured that there will be no
gravitational anomalies \fuji, and
we obtain the following expression for the Weyl anomaly
\eqn\anomaly{
\int d^n x {\sqrt g} \langle T \rangle \s
= -\hbar {\rm Tr } ( \s)
= -\hbar \lim_{m \to \infty} {\rm Tr} \biggl ( \s
e^{ {\nabla^2 +\xi R} \over m^2}\biggr )}
where
 the regulator, obtained  according to refs. \peter,
 is simply
given by the kinetic operator
$ {\cal R} = \nabla^2 + \xi R $.

As anticipated in the introduction, we would like to compute the anomaly
by  representing the trace on the right hand side of \anomaly\
as a trace in the Hilbert space of a quantum mechanical
system. Therefore, we should try to identify a system
with action $S[q]$ which has
the operator $ H = - (\nabla^2 + \xi R) $ as quantum Hamiltonian, so that we
can use path integral methods to compute the trace. The latter would then
read as follows
\eqn\repres{ {\rm Tr} \biggl ( \s e^{-\beta H } \biggr ) =
\int_{PBC} ({\cal D}q)\  \s \exp ( - S[q])}
where  the euclidean time $\beta $ is identified with
$m^{-2}$ and $PBC$ refers
to the periodic boundary condition $ q(0) = q(\beta)$
\ref\new{R.P. Feynman,
{\sl ``Statistical Mechanics''}, Benjamin, New York, 1972.}.
The correct system which does the job is that of  a particle moving in a
curved space.
To see this, express the Laplacian $\nabla^2$ by using the momentum
operator $\hat p_\m = -i g^{-{1\over 4}}
{\pa \over \pa q^\m}g^{{1\over 4}}$
\eqn\laplacian{-\nabla^2 =  g^{-{1\over 4}}\hat p_\m
{\sqrt g} g^{\m\n} \hat p_\n g^{-{1\over 4}}}
where $g = {\rm det} g_{\m\n}(q)$.
Note that $g_{\m\n}(q)$ is a function of the coordinates $q^\m$ (for notational
convenience, we sometimes do not indicate such a functional dependence).
In the classical limit, when the $p$'s commute with the $q$'s,
it corresponds to the classical Hamiltonian
\eqn\ham{ H_{cl} =g^{\m\n} p_\m p_\n}
which can be obtained from the configuration space Lagrangian
\eqn\lag{ L = g_{\m\n} \dot q^\m \dot q^\n }
where $\dot q^\m = {d \over dt } q^\m$. This Lagrangian describes a particle
with coordinates $q^\m$ moving in a manifold  with metric $g_{\m\n}(q)$.
We dedicate the next section to the analysis of the  path integral
quantization of such a system.

\newsec{The covariant path integral}

In this section, we want to consider the quantization
of the particle in curved spaces using a manifestly covariant path integral.
The object of interest is the transition amplitude
$\langle q^\m_f,t_f | q^\m_i,t_i \rangle$
for the particle starting
at point $q^\m_i$ at time $t_i$ and reaching point $q^\m_f$ at time $t_f$.
This is given by the covariant path integral\footnote{$^{\dagger}$}{After
Wick rotation to Euclidean time.}
\eqn\cpi{\eqalign{
&\langle q^\m_f,t_f | q^\m_i,t_i \rangle  =
\int_{q(t_i) = q_i}^{q(t_f)=q_f} (\tilde {\cal D} q)
\ \exp \biggl [ - {1\over \hbar} S[q]
\biggr ] \cr
&S[q] = \int_{t_i}^{t_f} dt\ \biggl (
{1\over 2} g_{\m\n} \dot q^\m \dot q^\n + v_\m \dot q^\m + s \biggr )\cr
& (\tilde {\cal D} q)
= \prod_{t_t< t < t_f} {\sqrt{ g(q(t))}} d^n q(t)\cr}}
where we have included in the action a vector potential $v_\m$ and a scalar
potential $s$ to be as general as possible.
This path integral is manifestly covariant since it is built from
manifestly covariant  objects.
 One can give a derivation of it by starting from the corresponding
phase space path integral and showing that the integration
over the momenta explicitly produces the
$\sqrt g$ factors in the measure of \cpi (see e.g. \ref\Abers{E.S. Abers and
B.W. Lee, {\sl Phys. Rep.} {\bf 9} (1973) 1.}).
However, the path integral in \cpi\ is so natural that one may simply take it
as a definition of the quantum theory. As it stands, it is still a formal
expression and one must tell how to
compute it, i.e. how to dicretize it. It is known that different prescriptions
for discretizing the path integral
correspond to different orderings of the $p$'s and $q$'s in the quantum
Hamiltonian.
We will compute the path integral by making a mode expansion of the histories
$q^\m(t)$, and eventually recognize the quantum Hamiltonian
by deriving the Schr\"odinger equation\footnote{$^{\dagger \dagger}$}
{More properly
we should call it
the diffusion equation, since we  use the Euclidean path integral.}
satisfied by the transition amplitude
$\langle q^\m_f,t_f | q^\m_i,t_i \rangle$.
We would like to compute the path integral
 perturbatively, in a loop expansion.
Unfortunately, the form in \cpi\ is not directly suitable for such an expansion
since the measure $(\tilde {\cal D} q)$ is not translational invariant
and the usual methods for deriving the propagators are  not applicable.
To overcome this difficulty, we introduce two anticommuting
ghost fields, $b$ and $c$,
and exponentiate the unwanted piece sitting in the measure.
We get
\eqn\ncpi{\langle q^\m_f,t_f | q^\m_i,t_i \rangle =
\int_{q(t_i) = q_i}^{q(t_f)=q_f} ({\cal D}q)({\cal D}b)({\cal D}c)
\ \exp \biggl [ - {1\over \hbar} (S[q] + S_{gh}[b,c,q])
\biggr ] }
where now the measures are translational invariant, i.e.
\eqn\measure{
({\cal D}q) = \prod_{t_t< t < t_f} d^n q(t),\ \ \
({\cal D}b)  = \prod_{t_t< t < t_f} d b(t),\ \ \
({\cal D}c) = \prod_{t_t< t < t_f} d c(t)}
 and the ghost action
is given by
\eqn\gh{S_{gh}[b,c,q] = \int_{t_i}^{t_f}
dt\ b {\sqrt {g(q)}} c }
Such a reformulation of the  path integral is still manifestly
covariant, provided one transforms $b$ and $c$ as weight ${1\over 2} $
densities under change of coordinates.
Explicitly
\eqn\transf{\eqalign{&
q^\m \rightarrow q^\m{}' =q^\m{}'(q^\n) \cr
&b \rightarrow b' = b \biggl ({\pa q'\over \pa q}\biggr )^{1\over 2}\cr
&c \rightarrow c' = c \biggl ({\pa q'\over \pa q}\biggr )^{1\over 2}\cr}}
Under these rules
the ghost action $S_{gh}[b,c,q]$  as well as the full measure
$({\cal D}q)({\cal D}b)({\cal D}c)$ are invariant objects.

\newsec{Two loop expansion and the quantum Hamiltonian}
We come now to the task of    evaluating
in a perturbative expansion  the covariant path integral defined in the
previous section. We will set $\hbar =1$ in the following.
The easiest way to proceed
is to make use of the general covariance of the path integral by
choosing Riemann normal coordinates (see e.g.
\ref\alv{L.  Alvarez-Gaum\'e, D.Z. Freedman and S. Mukhi, {\sl Ann. of Phys.}
{\bf 134} (1981) 85\semi
P.S Howe, G. Papadopulos and K.S. Stelle, {\sl Nucl. Phys.} {\bf B296}
(1988) 26.}).
We will use  Riemann normal coordinates $ z^\m$ centered at the
final point $ q^\m_f$.
Then, by definition, the Riemann normal coordinates
of a neighbouring point $q^\m$ are the components  of the tangent vector
at $q^\m_f$ of the geodesic joining $q^\m_f$ to $q^\m$ in unit time.
In particular  $q^\m_f$ has Riemann normal coordinates
 $z_f^\m = 0$.
Thus we see that the Riemann normal coordinates have a clear geometrical
meaning, namely they are tangent vectors belonging
to the tangent space of the manifold at the point $q_f^\m$.
This property will be  useful for
recognizing  the various geometrical quantities in the
 perturbative expansion
of the transition amplitude \ncpi.
Let's denote  the Riemann normal coordinates of
the initial point $q^\m_i$ in \ncpi\ by $x^\m$ and set $ \b = t_f - t_i$.
Then the geodesic
\eqn\geo{ z^\m_{cl} = - x^\m \tau;\ \ \ \ \ \ \tau = { t- t_f \over \b}}
is the classical path for vanishing  potentials $v_\m$ and $s$.
Before proceeding further, let us switch to the dimensionless time
variable $ \tau = {t - t_f \over \b}$, so that
the action reads as follows (with $ \dot q^\m = { d \over d \tau} q^\m $)
\eqn\4{ S = {1\over  \b } \int_{-1}^0 d\tau
\biggl ( {1 \over 2} g_{\m\n}
\dot q^\m \dot q^\n + \b  v_\m \dot q^\m + \b^2 s\biggr )}
and remark that to obtain the Schr\"odinger equation we need
terms up to $\b^2$ in the transition amplitude \ncpi\ (see our derivation
in eq. (4.21)). This approximation
 corresponds to a two loop expansion with $\b$ the loop counting
parameter.

In order to efficiently  carry out
the Riemann normal coordinates expansion
of the action in \4, it is useful to consider the field
$ q^\m(\tau, \l) $, defined for each $\tau$ to be the geodesic
connecting $q_f^\m$ at   $\l =0$ to
$ q^\m(\tau)$ at $\l=1$. Then by construction $q^\m (\tau,\l)$ satisfies
\eqn\1{{D \over d \l}{d q^\m  \over d \l}
\equiv {d^2 q^\m \over d \l^2} +
\Gamma^\m_{\n\r} {d q^\n \over d \l}  {d q^\r \over d \l}=0 }
and the Riemann normal coordinates of $q^\m(\tau)$ are given by
\eqn\rnc{ z^\m =  {d q^\m  \over d \l} \bigg \vert_{\l = 0}}
 The expansion of $S$ is  carried out by noticing that
\eqn\expa{ S [q(\tau)] =  S[q(\tau, \l=1)] =
\sum_{n=0}^{\infty} {1\over n!} {d^n S[q]  \over d \l^n}
\bigg \vert_{\l=0}=
\sum_{n=0}^{\infty} {1\over n!} {D^n S[q]  \over d \l^n}
\bigg \vert_{\l=0}}
so that one can proceed using covariant derivatives as well as
  the following identities
\eqn\prop{
{D z^\m \over d \l} = {D g_{\m\n}\over d \l} =0;
\ \ \ \
{D \dot q^\m \over d
\l} = {D z^\m \over d \tau};
\ \ \ \
\biggl [ {D \over d \l} , {D \over d \tau} \biggr ] V^\m =
z^\n
\dot q^\s R_{\n\s}{}^\m{}_\r V^\r}
where in the last equation $V^\m$ is an arbitrary vector
and  $z^\m = {d \over d\l} q^\m $
for any $\lambda$.

We obtain the following Riemann normal coordinates expansion
of the action in \4\
\eqn\2{ \eqalign{ S = {1\over \b} \int_{-1}^0 d \tau &\biggl [
{1\over 2} g_{\m\n}(0) \dot z^\m \dot z^\n +{1\over 6} R_{\m\n\r\s}(0)
z^\m \dot z^\n \dot z^\r z^\s \cr & + \b \biggl (
v_\m (0) \dot z^\m +  \nabla_\m v_\n(0) z^\m \dot z^\n
\biggr )
+ \b^2 s(0) +\cdots \biggr ]\cr }}
where we kept only terms contributing to two loops.
Next, we expand  a general path $z^\m$
around the classical solution as
\eqn\3{\eqalign{ & z^\m = z^\m_{cl} + y^\m \cr
& y^\m = \sum_{n=1}^\infty y^\m_n \sin (n \pi \tau) \cr}}
with $z^\m_{cl}$ given in \geo\ and
 where $y^\m$ is the quantum variable satisfying the boundary conditions
$y^\m (0) =y^\m(-1) =0$.
Inserting  this expansion in the
 quadratic leading  piece  of \2,
we obtain
\eqn\6{ S^{(2)} \equiv {1\over \b} \int_{-1}^0
 d\tau \ {1\over 2} g_{\m\n} (0) \dot z^\m \dot z^\n =
{1\over 2\b} g_{\m\n} (0) x^\m x^\n + {\pi^2 \over 4\b} g_{\m\n} (0)
\sum_{n=1}^\infty n^2 y^\m_n y^\n_n}
This leads to the  propagators
for the Fourier coefficients of the quantum field $y^\m$
\eqn\prop{ \langle y^\m_m y^\n_n \rangle = {2 \b \over  \pi^2 n^2}
\delta_{m,n} g^{\m\n}(0)}
as well as to the main normalization of the measure
\eqn\norm{
({\cal D} z) = A \prod_{m=1}^{\infty}
{\sqrt {g(0)}}
\prod_{\m=1}^n \biggl ( {\pi m^2 \over 4 \b} \biggr )^{1\over 2}
d y_m^\m}
with $A$ an yet unfixed normalization factor
(we have required here that apart from the normalization factor $A$,
the gaussian integrals over each Fourier mode give unity; the full
normalization will be derived later on, in the derivation of the Schr\"odinger
equation in eq. (4.21)).
Using the above propagator and treating the remaining part of
\2\ as a perturbation, we get a first contribution
to the transition amplitude\footnote{$^{\dagger}$}{We use
time translational invariance to set $t_f =0$.}
\eqn\cont{ \eqalign{ \Delta_1 \langle 0,0 | x^\m ,- \b \rangle =
& A\ \exp \biggl (
-{1\over 2 \b} g_{\m\n}(0) x^\m x^\n \biggr ) \biggl(
1 - \Big ( {1\over 18} R_{\m\n}(0) x^\m x^\n +
{ \b \over 36}  R(0) \Big ) M
\cr &-
 {5\over 72} R_{\m\n}(0) x^\m x^\n +
{ \b \over 36}  R(0)
+ v_\m(0) x^\m
+ {1\over 2} v_\m (0) v_\n (0) x^\m x^\n
\cr &
+{1\over 2} \nabla_\m v_\n (0) x^\m  x^\n
-\b s(o)
\biggr )\cr}}
where a divergent piece has been regulated by truncating the mode expansion
at the $M^{th}$ mode, while in the converging part the sum
$\sum_{n=1}^\infty {1\over n^2} = {\pi^2 \over 6}$
has been used.
In evaluating \cont\ we have kept only terms up to order $\b$
or terms which potentially can generate  pieces of order $\b$
when an integration over  the initial point $x^\m$ is performed.
The highest loop graph contributing to \cont\ is a two loop one.

Next, we look at  the ghost piece.
The expansion of the metric in Riemann normal coordinates is read off from
\2\
\eqn\aa{ g_{\m\n}(z) = g_{\m\n}(0) +{1\over 3} R_{\s(\m\n)\r} z^\s z^\r
+\cdots }
where $(\dots )$ denotes symmetrization in the enclosed indices.
Inserting this in the ghost action we get
\eqn\ghostaction{
S_{gh} = \int_{t_i}^{t_f} dt\ b {\sqrt {g(q) \over g(0)}} c =
\b \int_{-1}^0 d\tau \ b (1 + {1\over 6} R_{\m\n}(0) z^\m z^\n + \cdots ) c  }
Note that, for convenience, we left a $z^\m$-independent factor $\sqrt{g(0)}$
in the measure \norm.
At this point we have to discuss in more details the boundary conditions
on the ghost fields $b,\ c$. Since we  need to reproduce the
Jacobian in the measure \cpi, which excludes the end points $t_f$ and $t_i$,
we should not integrate on the boundary values of
$b$ and $c$. Therefore we require the following boundary
condition
for the ghost fields: $b(t_f) =b(t_i) =c(t_f) = c(t_f) =0$,
which is the
only boundary condition consistent with the algebraic character
of the  ghost  action  (i.e., it is the only boundary condition
consistent with the classical ghost field equations).
Taking this into account, we use the  mode decomposition
\eqn\mode{ b(\tau) = \sum_{n=1}^\infty b_n \sin (n \pi \tau); \ \ \ \
c(\tau) = \sum_{n=1}^\infty c_n \sin (n \pi \tau)}
and derive from the quadratic piece of the ghost action the
ghost propagator
\eqn\ghostprop{
\langle b_n c_m \rangle = -{2  \over \b} \d_{n,m} }
together with the correct normalization of the ghost measure
\eqn\norm{
({\cal D} b)({\cal D} c) =\prod_{m=1}^\infty
{2 \hbar \over \b }  d b_m d c_m }
Then, the interacting term in \ghostaction\  gives the remaining
perturbative contribution to the transition amplitude
\eqn\contribu{\eqalign{
\Delta_2 \langle 0,0 | x^\m ,-\b \rangle = &A\
\exp \biggl ( -{1\over 2 \b} g_{\m\n}(0)
x^\m x^\n \biggr )  \biggl (
1 + \Big ( {1\over 18} R_{\m\n}(0) x^\m x^\n +
{ \b \over 36}  R(0) \Big )
M
\cr
& - {1\over 72 } R_{\m\n} (0) x^\m x^\n
+{\b \over 72} R(0)
\biggr)\cr}}
Summing \cont\ with \contribu, we see that the potentially divergent terms
cancel and we are left with a finite expression
for the transition amplitude in the two loop approximation
\eqn\final{\eqalign{
\langle 0,0 | x^\m ,-\b\rangle = &A\
  \exp \biggl ( -{1\over 2 \b} g_{\m\n}(0)
x^\m x^\n \biggr )  \biggl (
1-
 {1\over 12} R_{\m\n}(0) x^\m x^\n +
{ \b \over 24}  R(0)
\cr &+ v_\m(0) x^\m
+ {1\over 2 } v_\m (0) v_\n (0) x^\m x^\n
+{1\over 2 } \nabla_\m v_\n (0) x^\m  x^\n
-\b s(o)
\biggr )\cr}}
It is worth to note that we have performed the full path integral even
in the limit $\b \to 0$. Approximating the transition amplitude
by considering only the contribution coming from the extremal classical
trajectory is incorrect when the space is curved.

We derive now the Schr\"odinger equation to recognize the quantum
Hamiltonian associated with the
transition amplitude
\final\  and fix the  coefficient $A$ in it.
The transition amplitude is used to evolve
the wave function $\Psi(q^\m,t)$ as follows
\eqn\wave{
\Psi (q^\m_f,t_f)
= \int d^n q_i \sqrt{g(q_i)}\  \langle q^\m_f, t_f | q^\m_i,t_i
\rangle \Psi (q^\m_i, t_i) }
Using \wave\  in Riemann normal coordinates
and Taylor expanding the wave function as well as the measure
on the right hand side,
we get
\eqn\wave{\eqalign{
\Psi (0,0)
&= \int d^n x \sqrt{g(0)} \  (1 +{1 \over 6} R_{\m\n}(0) x^\m x^\n +\cdots)
   \langle 0,0 | x^\m, -\b
\rangle \cr &\times
(\Psi (0, 0) - \b \dot \Psi (0,0) + x^\m {\pa \Psi \over \pa x^\m}
(0,0)
+ {1\over 2} x^\m x^\n
{\pa^2 \Psi \over \pa x^\m \pa x^\n}(0,0)  +\cdots) \cr}}
and inserting \final\ in \wave, we obtain as a  consistency requirement
\eqn\A{ A=  (2 \pi \b )^{- {n\over 2}} }
as well as the Euclidean  Schr\"odinger equation (diffusion equation)
\eqn\Sc{ \eqalign { &- \dot \psi =  H \psi
 \cr & H = -{1\over 2} \nabla^2 - v^\m \pa_\m
- {1\over 2} v_\m v^\m - {1\over 2} ( \nabla_\m v^\m)
- {1 \over 8}R
+ s \cr}}
This is the correct quantum Hamiltonian
associated with the path integrals \cpi\ and \ncpi.
It can be rewritten in the following manifestly gauge invariant
way
\eqn\gaugeinv{ H = -{1 \over 2}(\nabla^\m + v^\m)(\partial_\m + v_\m)
- {1 \over 8}R
+ s }

Now, as promised, we turn to the computation of the Weyl anomaly in two
dimensions.
We proceed in full generality and
 make use of   \final\
and \Sc\ to derive a lemma, already presented in \ref\me{
F. Bastianelli, P. van Nieuwenhuizen and A. Van Proeyen, {\sl Phys. Lett.}
{\bf B253} (1991) 67.},
for  computing general algebraic anomalies. Namely, we want to compute
\eqn\lemma{ \eqalign{& I = \lim_{m \to \infty}{\rm Tr} (\s e^{{\cal R}\over
m^2}
   )
\cr & {\cal R } = \nabla^2 + V^\m \pa_\m +S \cr}}
which corresponds to
\eqn\bb{I = \lim_{\b \to 0}
\int_{PBC} (\tilde {\cal D} q)
\exp (- S[q])}
 where $PBC$ stands for the periodic
boundary conditions $ q^\m (0) =q^\m(-\b)$,
and where, according to \Sc, we have to use the potentials
\eqn\pot{\eqalign{ &s=- {1\over 2} S +{1\over 8} R +{1\over 2}
v_\m v^\m
+{1\over 2} \nabla_\m v^\m \cr
&v_\m =  {1\over 2} V_\m
\cr}}
to recover $\cal R$ as $-2$ times the quantum Hamiltonian.
Actually, in the limit $\b \rightarrow 0$ there are also
divergent pieces in \bb\
which should be discarded. This corresponds to the fact that in computing the
effective action in quantum field theory, one renormalizes away all the
divergent pieces. In particular, anomalies are always finite.
Thus, one should interpret the ``$\lim_{\b \to 0} $''
symbol as  meaning ``pick the $\b$ independent part of''.
With these cautionary words for interpreting our formulae,
we get
\eqn\last{ \eqalign{ I& = \lim_{\b \to 0}
\int d^2q {\sqrt{ g(q)}}\
\s (q) \langle q^\m, 0 | q^\m, -\b \rangle
\cr & =\int {d^2 q
\over 4 \pi }{\sqrt{ g(q)}} \
\s(q) \biggl ( S - {1\over 2} \nabla_\m V^\m - {1\over 4} V_\m V^\m
-{1\over 6} R \biggr ) \cr}}
In particular, for vanishing $S$ and $V^\m$, we get the
well-known result for the Weyl anomaly in \anomaly, specialized
to the case of a scalar field
in two dimensions
\eqn\cc{ \langle T \rangle =
{ \hbar \over 24 \pi} R }

\newsec{Conclusions}
We have shown that Weyl anomalies can be computed by using a
quantum mechanical representation of the ``Fujikawa'' Jacobian.
{}From the normalization \A\ and the formula \anomaly, we can see
that
the Weyl anomaly in $n$ dimensions is given by the $\b^{n \over 2}$
perturbative contribution to the quantum mechanical
path integral, indicating that in odd
dimensions the Weyl anomaly always vanish, since fractional powers
of $\b$ never arise in the quantum mechanical loop expansion.
In even dimensions  $n=2k$,
the terms contributing to the anomaly contain connected graphs up to
$k+1$ loops. We have explicitly computed the Weyl anomaly
for a scalar field in two dimension.
This result has been achieved by using a novel reformulation
of the path integral for a particle
in curved spaces which makes use of ghost fields. The advantage
of using ghost fields comes from the fact that one can
obtain translational invariant path integral measures, suitable for
generating the perturbative expansion.
Our approach  has been rather pragmatical
in that we used the quantum mechanics as a mathematical tool to
compute an infinitesimal path integral Jacobian, as pioneered
in the computation of chiral anomalies \ag, where supersymmetric
quantum mechanics was used.
{}From this perspective, one  does not care about the nature
of the particular  quantum mechanical system used to represent the required
trace. Typically, one considers the quantum mechanical Hilbert space
to be of a fictitious, non-physical nature \dewitt.
However, at a closer inspection, one can reinterpret our computation
as being the evaluation of the Weyl anomaly  using  the first
quantized  description of a scalar field coupled to gravity.
The quantum mechanical Hilbert space is then not at all fictitious,
but it is directly related to the Hilbert space of the
first quantized Klein-Gordon field.
This interpretation can in fact be extended to the  Schwinger-De Witt
method for calculating the Klein-Gordon propagator in curved spaces \dewitt.
In that method, the Klein-Gordon  propagator is obtained by integrating
over the proper time a kernel which satisfies the heat equation,
and which, in fact, is given by
our equation \final.
The proper time can be identified as the
integral over the moduli space of the one dimensional metric, which enters the
action of the first quantized scalar field. Such an action would reduce
to equation \4\ upon
gauge fixing and by ignoring the corresponding ghost fields,
 which however do not
couple to the target-space geometry and which should not be confused
with the  ghosts introduced in section 3.

\bigbreak \bigskip\bigskip
{\bf Acknowledgements}

It is a pleasure to thank Peter van Nieuwenhuizen, who has prompted my
interest in the subject.

\listrefs
\end